\documentclass[aps,prb,twocolumn]{revtex4}
\usepackage{graphicx}
\usepackage{dcolumn}
\usepackage{amsmath}
\usepackage{amssymb}
\usepackage{subfigure,amsmath,verbatim,moreverb,bm}
\def\be{\begin{equation}}
\def\ee{\end{equation}}
\def\ber{\begin{eqnarray}}
\def\eer{\end{eqnarray}}

\def\rv{{\bf r}}

\def\fv{{\bf f}}

\begin{document}
\title{Kohn-Sham density functional theory for quantum wires in arbitrary
correlation regimes}
\author{Francesc Malet,$^1$ Andr\'e Mirtschink,$^1$  Jonas C. Cremon,$^2$  Stephanie M. Reimann,$^2$ and Paola Gori-Giorgi$^1$}
\affiliation{$^1$Department of Theoretical Chemistry and Amsterdam Center for 
Multiscale Modeling, 
FEW, Vrije Universiteit, De Boelelaan 1083, 1081HV Amsterdam, The Netherlands\\
$^2$Mathematical Physics, Lund University, LTH, P.O. Box 118, SE-22100 Lund, Sweden}

\date{\today}

\begin{abstract}
We use the exact strong-interaction limit of the Hohenberg-Kohn energy density 
functional to construct an approximation for the exchange-correlation term of
the Kohn-Sham approach. The resulting exchange-correlation potential is able 
to capture the features of the strongly-correlated regime without breaking 
the spin or any other symmetry. In particular, it shows ``bumps'' (or barriers) 
that give rise to charge localization at low densities and that are a 
well-known key feature of the exact Kohn-Sham potential for strongly-correlated systems.
Here we illustrate this approach for the study of both weakly and strongly correlated 
model quantum wires, comparing our results with those obtained with the configuration 
interaction method and with the usual Kohn-Sham local density approximation.
\end{abstract}
\maketitle

\section{Introduction}

In semiconductor nanostructures, the regime of strong correlation is reached 
when the electronic density becomes low enough so that the Coulomb repulsion 
becomes dominant with respect to the kinetic energy of the electrons. From the
purely fundamental point of view, the study of the strongly-interacting limit 
in such systems is interesting since charge localization, reminiscent of the 
Wigner crystallization \cite{Wig-PR-34} of the bulk electron gas, is expected to 
occur at low densities. 

A lot of previous theoretical work on Wigner 
localization in nanostructures has focused on finite-sized quantum dots (see,
for example, Refs. \onlinecite{CreHauJefSar-PRB-99,EggHauMakGra-PRL-99,
YanLan-PRL-99, ReiKosMan-PRB-00,FilBonLoz-PRL-01,GhoGucUmrUllBar-PRB-07}), 
and the crossover from liquid to localized states in the transport properties 
of the nanostructure has been addressed. \cite{CavGioSasKra-NJP-09,
EllIhnYanLanEnsDriGos-PRL-06} In quasi one-dimensional nanosystems, signatures 
of Wigner localization were observed experimentally in one-dimensional 
cleaved-edge overgrowth structures, \cite{AusSteYacTseHalBalPfeWes-Sci-05} 
or in the transport properties of InSb nanowire quantum-dot systems. 
\cite{KriCreNilXuSamLinWacRei-PRB-11} More recent experimental work clearly 
identified the formation of Wigner molecules in a one-dimensional quantum 
dot that was capacitively coupled to an atomic force microscope 
probe. \cite{ZiaCavSas-PRB-12} Wigner localization has also been investigated 
in other 1D systems such as carbon nanotubes. \cite{DesBoc-NP-08,SecRon-PRB-10,
SecRon-PRB-12} (For a review, see Ref. \onlinecite{DesBocGlaYac-Nat-10}).
Finally, regarding practical applications, Wigner-localized systems have been 
shown to be potentially useful,{\it e.g.}, for quantum-computing 
purposes.\cite{DesBoc-NP-08,TayCal-PRA-08}   

When trying to model electronic strongly-correlated systems, however, the 
commonly employed methodologies encounter serious difficulties of different 
nature. On the one hand, the configuration 
interaction (CI) approach, despite being in principle capable of describing any 
correlation regime, is in practice limited to the study of small systems 
with only very few particles due to its high computational cost, which scales 
exponentially with the number of particles, $N$. Such numerical difficulties 
get even worse in the very strongly-correlated limit due to the degeneracy of the 
different quantum states and the consequent need of considering larger Hilbert 
spaces in the calculations. Other wave-function methods like Quantum Monte Carlo\cite{GhoGucUmrUllBar-NP-06,GhoGucUmrUllBar-PRB-07,GucGhoUmrBar-PRB-08} (QMC) and density matrix renormalization group (DMRG),\cite{StoWagWhiBur-PRL-12} which rely to some extent on various approximations, can treat systems larger than the CI approach, but are still computationally expensive and limited to $N\lesssim 100$. 

The much cheaper Kohn-Sham (KS) density functional theory (DFT),\cite{HohKoh-PR-64,KohSha-PR-65}  
which allows to treat thousands of electrons, is the method of choice to treat larger 
quantum systems. However, all the currently available approximations for the 
exchange-correlation functional fail to describe the strongly-correlated 
regime\cite{AniZaaAnd-PRB-91,GruGriBae-JCP-03,CohMorYan-SCI-08,GhoGucUmrUllBar-PRB-07,BorTorKosManAbeRei-IJQC-05,AbePolXiaTos-EJPB-07} even at the qualitative level.  Allowing spin- and spatial-symmetry breaking 
may yield reasonable total energies, without, however, capturing the physics of 
charge localization in non-magnetic systems. Moreover, broken symmetry solutions 
often yield a wrong characterization of various properties and 
the rigorous KS DFT framework is partially lost (see, {\it e.g.}, 
Refs.~\onlinecite{AniZaaAnd-PRB-91,BorTorKosManAbeRei-IJQC-05,StoWagWhiBur-PRL-12}). 
 
KS DFT is, in principle, an exact theory that should be able to yield the exact energy 
and density even in the case of strong electronic correlation, without artificially 
breaking any symmetry. However, when dealing with practical KS DFT, one could expect 
that the non-interacting reference system introduced by Kohn and Sham might not be the 
best choice when trying to address systems in which the electron-electron interactions 
play a dominant role. For many 
years, huge efforts have been made in order to try to get a better 
characterization and understanding of the properties of the exact Kohn-Sham 
reference system (see e.g. Refs. \onlinecite{Ver-PRL-08,HelTokRub-JCP-09,
TemMarMai-JCTC-09,TeaCorHel-JCP-09,TeaCorHel-JCP-10,KurSteKhoVerGro-PRL-10,
SteKur-PRL-11,KarPriVer-PRL-11,StoWagWhiBur-PRL-12,BerLiuBurSta-PRL-12,
RamGod-PRL-12,BuiBaeSni-PRA-89,FilUmrTau-JCP-94,GrivanBae-PRA-95,GrivanBae-JCP-96,
ColSav-JCP-99,CohMorYan-SCI-08,MorCohYan-PRL-09,Vie-PRB-12}). All these
works reflected the large difficulties encountered when trying to obtain 
adequate approximations to describe strong correlation in the exact KS theory.\cite{CohMorYan-CR-12}

An alternative density-functional framework, based on the study of the strongly-interacting limit of the Hohenberg-Kohn density functional, was presented  
in Ref.~\onlinecite{GorSeiVig-PRL-09}. In this approach, a reference system with 
infinite correlation between the electrons was considered instead of the 
non-interacting one of Kohn and Sham. The two formalisms can therefore be seen 
as complementary to each other and, indeed, the first results obtained with this 
so-called {\em strictly-correlated-electrons} (SCE) DFT, presently limited to 
either 1D or spherically-symmetric systems, showed its ability to describe 
systems in the extreme strongly-correlated regime with a much better accuracy than standard
KS DFT.\cite{GorSeiVig-PRL-09,GorSei-PCCP-10} 
On the downside, however, SCE DFT requires that one knows {\em a priori} that 
the system is in the strong-interaction regime, and it fails as soon as the fermionic nature of the electrons plays a significant role. \cite{GorSei-PCCP-10} Furthermore, the formalism lacks 
some of the appealing properties of the Kohn-Sham approach, such as its capability to predict (at least in principle) exact ionization 
energies. Also, crucial concepts widely employed in 
solid state physics and in chemistry, such as the Kohn-Sham orbitals and orbital energies, are 
totally absent in SCE DFT.

Very recently, a new approach that combines the advantages of the KS and the SCE DFT formalisms, consisting in approximating the Kohn-Sham 
exchange-correlation energy functional with the strong-interaction limit of the Hohenberg-Kohn energy density 
functional, has been proposed.\cite{MalGor-PRL-12}  Pilot tests of this new ``KS SCE'' framework showed that it is able to 
capture the features of both the weakly and the strongly-correlated regimes in 
semiconductor quantum wires, as well as the so-called $2k_F\to 4 k_F$ crossover 
occurring in between them, while keeping (at least for 1D systems) a computational cost comparable to  the one of 
standard KS DFT with the local-density approximation (LDA). In other words, the SCE functional yields a highly 
non-local approximation for the exchange-correlation energy functional, which is able to capture key features 
of strong correlation within the KS scheme, without any artificial symmetry breaking.

The main purpose of this work is to further investigate this new KS SCE method, by discussing its exact formal properties and, for the prototypical case of (quasi)-1D quantum wires, by also performing full CI calculations to compare electronic densities, total energies and one-electron removal energies in different regimes of correlation. We find that the KS SCE results are qualitatively right at all correlation regimes, representing an important advance for KS DFT.   However, while one-electron removal energies are quite accurate, total energies and ground-state densities are still quantitatively not always satisfactory, and therefore we also discuss the construction of corrections to KS SCE. In particular, we investigate here a simple local correction, which, however, turns out to give rather disappointing results, suggesting that to further improve KS SCE we need semi-local or fully non-local density functionals.

 The paper is organized as follows. In the next Sec.~\ref{sec_methodology} we describe the KS SCE approach, illustrating and discussing its features beyond what was reported in Ref.~\onlinecite{MalGor-PRL-12}. In Sec.~\ref{sec_Q1D} we introduce the quasi-1D systems we have addressed, and in 
Sec.~\ref{sec_results} we present our results, comparing the performances of KS SCE with the ``exact'' CI results, with the standard KS LDA method, and discussing KS SCE with a simple local correction.
Finally, in Sec.~\ref{sec_conc} we draw some conclusions, as well as an outlook for future works. 

Hartree (effective) atomic units are used throughout the paper.

\section{Theory and Methodology}
\label{sec_methodology}

\subsection{KS and SCE DFT}

In the formulation of Hohenberg and Kohn \cite{HohKoh-PR-64} the ground-state density 
and energy of a many-electron system are obtained by minimizing the energy density functional
\be\label{EnergyFunctional}
E[\rho] = F[\rho]+\int d\rv\, v_{\rm ext}(\rv)\,\rho(\rv)
\ee
with respect to the density $\rho(\rv)$.
In Eq.~\eqref{EnergyFunctional} $v_{\rm ext}(\rv)$ is the external potential and $F[\rho]$ is a universal 
functional of the density, defined as the minimum of the internal energy 
(kinetic energy $\hat T$ plus electron-electron repulsion $\hat V_{ee}$) with 
respect to all the fermionic wave functions $\Psi$ that yield the 
density $\rho(\rv)$, \cite{Lev-PNAS-79}
\be
F[\rho]=\min_{\Psi\to\rho}\langle\Psi|\hat T+\hat V_{ee}|\Psi\rangle.
\label{eq_HK}
\ee
In order to capture the fermionic nature of the electronic density, Kohn and 
Sham \cite{KohSha-PR-65} introduced the functional $T_s[\rho]$ by minimizing 
the expectation value of  $\hat T$ alone over all the fermionic wave functions 
yielding the given $\rho({\bf r})$, \cite{Lev-PNAS-79}
\be
T_s[\rho]=\min_{\Psi\to\rho}\langle\Psi|\hat T|\Psi\rangle,
\label{eq_Ts}
\ee
thus introducing a reference system of non-interacting electrons with the 
same density as the physical, interacting, one. The remaining part
of $F[\rho]$, defining the Hartree and the exchange-correlation functionals, 
$F[\rho]-T_s[\rho]\equiv E_{\rm Hxc}[\rho]\equiv E_{\rm H}[\rho] + E_{\rm xc}[\rho]$,
is then approximated. The minimization of the total energy functional $E[\rho]$ 
with respect to the density yields the well-known single-particle Kohn-Sham 
equations \cite{KohSha-PR-65}
\be
\left(-\frac{1}{2}\nabla^2 + v_{\rm KS}[\rho](\rv)\right)\phi_i(\rv) = \varepsilon_i \phi_i(\rv) \; ,
\label{eq_KS}
\ee
where $v_{\rm KS}(\rv)\equiv v_{\rm ext}(\rv) + 
\delta E_{\rm Hxc}[\rho]/\delta \rho(\rv)\equiv v_{\rm ext}[\rho](\rv) +  v_H[\rho](\rv) +
v_{xc}[\rho](\rv)$ is the one-body local Kohn-Sham potential, 
with $v_H[\rho](\rv)$ and $v_{xc}[\rho](\rv)$ being, respectively, the Hartree 
and the exchange-correlation parts. The solutions $\phi_i$ of Eqs.~(\ref{eq_KS}) 
are the so-called Kohn-Sham orbitals, which yield the electronic density through 
the relation $\rho({\bf r}) = \sum_i|\phi_i({\bf r})|^2$, with the sum running only over 
occupied orbitals. Notice that here we work with the original, spin-restricted, KS 
scheme, in which we have the same KS potential for spin-up and spin-down electrons. 

The HK functional of Eq. (\ref{eq_HK}) and the KS functional of Eq.~(\ref{eq_Ts}) 
can be seen as the particular values at $\lambda=1$ and at $\lambda=0$ of a more general 
functional $F_\lambda[\rho]$ in which the coupling-strength interaction is rescaled 
with a real parameter $\lambda$, {\it i.e.},
\be
F_\lambda[\rho]=\min_{\Psi\to\rho}\langle\Psi|\hat T+\lambda\hat V_{ee}|\Psi\rangle.
\label{eq_HKlambda}
\ee
A well-known exact formula for the Hartree-exchange-correlation functional
$E_{\rm Hxc}[\rho]$ is \cite{LanPer-SSC-75, GunLun-PRB-76}
\be
E_{\rm Hxc}[\rho]=\int_0^1 \langle\Psi_\lambda[\rho]|V_{ee} |\Psi_\lambda[\rho]\rangle\, 
d\lambda\equiv \int_0^1 V_{ee}^{\lambda}[\rho]\, d\lambda,
\label{eq_Ehxc}
\ee
where $\Psi_\lambda[\rho]$ is the minimizing wave function in Eq. (\ref{eq_HKlambda}).

In the strictly-correlated-electrons DFT (SCE DFT) formalism, one considers the
strong-interaction limit of the Hohenberg-Kohn functional, $\lambda\to\infty$,
which corresponds to the functional \cite{Sei-PRA-99,SeiPerLev-PRA-99,SeiPerKur-PRL-00,SeiGorSav-PRA-07}
\be
V_{ee}^{\rm SCE}[\rho]\equiv\min_{\Psi\to\rho}\langle\Psi|\hat V_{ee}|\Psi\rangle,
\label{eq_VeeSCE}
\ee
i.e., the minimum of the electronic interaction alone over all the wave functions 
yielding the given density $\rho({\bf r})$. This limit has been first studied in the 
seminal work of Seidl and coworkers \cite{Sei-PRA-99,SeiPerLev-PRA-99,SeiPerKur-PRL-00}, 
and later formalized and evaluated exactly in a rigorous mathematical way in 
Refs.~\onlinecite{SeiGorSav-PRA-07,GorSei-PCCP-10,RasSeiGor-PRB-11,ButDepGor-PRA-12}. 
The functional $V_{ee}^{\rm SCE}[\rho]$ also defines a reference system complementary to the non-interacting 
one of the Kohn-Sham kinetic energy $T_s[\rho]$, one composed by infinitely-correlated electrons, with 
zero kinetic energy. This implies that, analogously as in a set of confined classical 
repulsive charges, which arrange themselves seeking for the stable 
spatial configuration that minimizes their interaction energy, in the SCE reference 
system the position of one electron uniquely determines the position of the remaining 
ones, always under the constraint imposed by Eq. \eqref{eq_VeeSCE} that the density 
at each point is equal to that of the quantum-mechanical system with $\lambda=1$, $\rho({\bf r})$.

More precisely, the functional $V_{ee}^{\rm SCE}[\rho]$ is constructed\cite{SeiGorSav-PRA-07} by 
considering that the admissible configurations of $N$ electrons in $d$ dimensions 
are restricted to a $d-$dimensional subspace $\Omega_0$ of the full classical $Nd-$dimensional 
configuration space. A generic point of $\Omega_0$ has the form
\be
{\bf R}_{\Omega_0}({\bf s})=({\bf f}_1({\bf s}),.....,{\bf f}_N({\bf s})),
\label{R_Omega_0}
\ee
where ${\bf s}$ is a $d$-dimensional vector that determines the position of, say,
electron ``1'', and ${\bf f}_i({\bf s})$ $(i=1,...,N)$, 
with ${\bf f}_1({\bf s})={\bf s}$, are the so-called {\em co-motion functions}, 
which determine the position of the $i$-th electron as a function of ${\bf s}$. 
The co-motion functions are implicit non-local functionals of the given density 
$\rho({\bf r})$, \cite{SeiGorSav-PRA-07,GorVigSei-JCTC-09,GorSeiVig-PRL-09,ButDepGor-PRA-12} 
and solution of a set of differential equations that ensure the invariance of $\rho$ 
under the coordinate transformation ${\bf s}\to{\bf f}_i({\bf s})$, i.e., 
\be
\rho({\bf f}_i({\bf s}))d{\bf f}_i({\bf s}) = \rho({\bf s})d{\bf s},
\label{eq_fi}
\ee
or, equivalently, that the probability of finding the electron $i$ at 
${\bf f}_i({\bf s})$ is equal to that of finding the electron ``1'' at 
${\bf s}$. At the same time, the ${\bf f}_i({\bf s})$ must satisfy group properties that 
ensure the indistinguishability of the $N$ electrons.\cite{SeiGorSav-PRA-07,ButDepGor-PRA-12} 

The functional $V_{ee}^{\rm SCE}[\rho]$ can then be written in 
terms of the co-motion functions ${\bf f}_i$ as \cite{SeiGorSav-PRA-07,MirSeiGor-JCTC-12}
\ber
V_{ee}^{\rm SCE}[\rho]&=&\int d{\bf s}\frac{\rho({\bf s})}{N}\sum_{i= 1}^{N-1}
\sum_{j= i+1}^N\frac{1}{|{\bf f}_i({\bf s})-{\bf f}_j({\bf s})|} 
\nonumber \\
&=&\frac{1}{2}\int d{\bf s}\,\rho({\bf s})\sum_{i= 2}^N
\frac{1}{|{\bf s}-{\bf f}_i({\bf s})|},
\label{eq_VeeSCE2}
\eer
just as $T_s[\rho]$ is written in terms of the Kohn-Sham 
orbitals $\phi_i(\rv)$. The equivalence of the two 
expressions for $V_{ee}^{\rm SCE}[\rho]$ in Eq.~\eqref{eq_VeeSCE2} has been proven in Ref.~\onlinecite{MirSeiGor-JCTC-12}.

Since in the SCE system the position of one electron determines all 
the other $N-1$ relative positions, the net repulsion felt by an electron at 
position ${\bf r}$ due to the other $N-1$ electrons becomes a function of ${\bf r}$ itself. For a given density
$\rho_0({\bf r})$, this effect can be {\it exactly} 
transformed \cite{SeiGorSav-PRA-07,GorSeiVig-PRL-09,ButDepGor-PRA-12} 
into a local one-body effective external potential 
$v_{\rm SCE}[\rho_0]({\bf r})$ that compensates the total 
Coulomb force on each electron when all the particles are at their 
respective positions ${\bf f}_i[\rho_0]({\bf r})$, i.e., such that \cite{SeiGorSav-PRA-07}
\be
\nabla v_{\rm SCE}[\rho_0](\rv)
=\sum_{i= 2}^N \frac{\rv-\fv_i[\rho_0](\rv)}{|\rv-\fv_i[\rho_0](\rv)|^3}.
\label{eq_vSCE}
\ee
In terms of the classical-charge analogue, $v_{\rm SCE}[\rho_0]({\bf r})$ 
can thus be seen as an external potential for which the total classical potential energy
\be
E_{\rm pot}(\rv_1,...\rv_N) \equiv \sum_{i= 1}^{N-1}\sum_{j= i+1}^N\frac{1}{|{\bf r}_i - {\bf r}_j|}
+ \sum_{i= 1}^N v_{\rm SCE}[\rho_0](\rv_i) 
\ee
is minimum when the electronic positions reside on the subset ${\bf R}_{\Omega_0}$,
i.e., when $\rv_i = \fv_i[\rho_0]({\bf r})$ or, equivalently, when the associated density
at each point is equal to $\rho_0({\bf r})$. For an arbitrary density $\rho({\bf r})$, 
the potential-energy density functional defined as
\be
E_{\rm pot}^{\rm SCE}[\rho]\equiv V_{ee}^{\rm SCE}[\rho] + 
\int v_{\rm SCE}[\rho_0](\rv) \rho(\rv) d\rv
\ee
will then satisfy the stationarity 
condition $\delta E_{\rm pot}^{\rm SCE}[\rho]/\delta \rho({\bf r})\big|_{\rho=\rho_0}=0$, i.e., 
we will have that
\be
\frac{\delta V_{ee}^{\rm SCE}[\rho]}{\delta \rho(\rv)}\bigg|_{\rho=\rho_0}=
-v_{\rm SCE}[\rho_0](\rv).
\label{eq_funcder}
\ee
Notice that Eq.~(\ref{eq_funcder}) involves the functional derivative of a 
highly non-local implicit functional of the density, defined by 
Eqs.~(\ref{eq_fi})-(\ref{eq_VeeSCE2}). This, however turns out to reduce to a local 
one-body potential that can be easily calculated from the integration 
of Eq.~(\ref{eq_vSCE}) once the co-motion functions are obtained via 
Eq.~(\ref{eq_fi}). This shortcut to compute the functional derivative 
of $V_{ee}^{\rm SCE}[\rho]$ is extremely powerful for including strong-correlation 
in the KS formalism.\cite{MalGor-PRL-12}

\subsection{Zeroth-order KS-SCE approach}

Equations~(\ref{eq_vSCE}) and (\ref{eq_funcder}) show how the effects of strong 
correlation, captured by the limit $\lambda\to\infty$ of $F_\lambda[\rho]$ and 
rigorously represented by the highly non-local functional $V_{ee}^{\rm SCE}[\rho]$, 
are {\em exactly} transferred into the one-body potential $v_{\rm SCE}[\rho]$. 
The KS SCE approach to zeroth order consists in using this property to approximate 
the Hartree-exchange-correlation term of the Kohn-Sham potential as
\be 
\frac{\delta E_{\rm Hxc}[\rho]}{\delta \rho(\rv)} \approx\tilde{v}_{\rm SCE}[\rho](\rv), \quad
 \tilde{v}_{\rm SCE}[\rho](\rv)\equiv-v_{\rm SCE}[\rho](\rv). 
\label{eq_vHxcvSCE}
\ee
Notice that we have defined 
$\tilde{v}_{\rm SCE}[\rho](\rv)=-v_{\rm SCE}[\rho](\rv)$, 
as here we seek an effective potential for KS theory, which corresponds to the net 
electron-electron repulsion acting on an electron at position $\rv$, while the
effective potential for the SCE system of Eq.~\eqref{eq_vSCE} {\it compensates} the 
net electron-electron repulsion.

More rigorously, by considering the $\lambda\to \infty$ expansion of the integrand
of Eq.~(\ref{eq_Ehxc}) one obtains \cite{Sei-PRA-99,SeiPerLev-PRA-99,SeiPerKur-PRL-00,SeiGorSav-PRA-07,GorVigSei-JCTC-09}
\be
V_{ee}^{\lambda\to \infty}[\rho]=V_{ee}^{\rm SCE}[\rho]+
\frac{V_{ee}^{\rm ZPE}[\rho]}{\sqrt{\lambda}}+O(\lambda^{-p}),
\label{eq_exp}
\ee
where the acronym ``ZPE'' stands for ``zero-point energy'', and 
$p\geq$5/4 -- see Ref.~\onlinecite{GorVigSei-JCTC-09} for further details.
By inserting the expansion of Eq.~(\ref{eq_exp}) into Eq.~(\ref{eq_Ehxc}) one
obtains an approximation for $E_{\rm Hxc}[\rho]$,
\be
E_{\rm Hxc}[\rho]\approx V_{ee}^{\rm SCE}[\rho] + 2\,V_{ee}^{\rm ZPE}[\rho] + ...
\label{eq_EHxcSCE}
\ee
We consider here only the first term, corresponding to a zeroth-order
expansion around $\lambda=\infty$, 
i.e., $E_{\rm Hxc}[\rho]\approx V_{ee}^{\rm SCE}[\rho]$, which yields 
Eq.~\eqref{eq_vHxcvSCE} for the corresponding functional derivatives. 

Taking into account the definition of the functional $V_{ee}^{\rm SCE}[\rho]$, 
Eq.~(\ref{eq_VeeSCE}), the zeroth-order KS SCE is equivalent to approximate 
the minimization over $\Psi$ in the HK functional of Eq.~(\ref{eq_HK}) as
\ber
\min_{\Psi\to\rho}\langle\Psi|\hat T + \hat V_{ee}|\Psi\rangle &\approx&
\min_{\Psi\to\rho}\langle\Psi|\hat T|\Psi\rangle +
\min_{\Psi\to\rho}\langle\Psi|\hat V_{ee}|\Psi\rangle 
\nonumber\\
&=& T_s[\rho] + V_{ee}^{\rm SCE}[\rho].
\label{eq_FKSSCE}
\eer
The KS SCE approach thus treats both the kinetic energy and the electron-electron 
repulsion on the same footing, combining the advantages of both KS and SCE DFT and 
therefore allowing one to address both the weakly- and the 
strongly-interacting regime, as well as the crossover between them.\cite{MalGor-PRL-12}
Indeed, from the scaling properties\cite{LevPer-PRA-85} of the functionals 
$F[\rho]$, $T_s[\rho]$ and $V_{ee}^{\rm SCE}[\rho]$ it derives that the approximation 
of Eq.~\eqref{eq_FKSSCE} becomes accurate both in the weak- and in the strong-interaction 
limits, while probably less precise in between.  
To use the scaling relations\cite{LevPer-PRA-85} one defines, for electrons in $D$ dimensions, 
a scaled density $$\rho_\gamma(\rv)\equiv\gamma^D\rho(\gamma\, \rv) \qquad \gamma > 0.$$
We then have\cite{LevPer-PRA-85,GorSei-PCCP-10} 
\begin{eqnarray}
T_s[\rho_\gamma] & = & \gamma^2\,T_s[\rho] \label{eq_scalTs} \\
V_{ee}^{\rm SCE}[\rho_\gamma] & = & \gamma\, V_{ee}^{\rm SCE}[\rho] \label{eq_scalVeeSCE} \\
F[\rho_\gamma]& = & \gamma^2\,F_{1/\gamma}[\rho], \label{eq_scalF}
\end{eqnarray}
where $F_{1/\gamma}[\rho]$ 
means\cite{LevPer-PRA-85} that the Coulomb coupling constant $\lambda$ in $F_\lambda[\rho]$ of 
Eq.~\eqref{eq_HKlambda} has been set equal to $1/\gamma$.
We then see that both sides of Eq.~\eqref{eq_FKSSCE} tend to $T_s[\rho_\gamma]$ when $\gamma\to\infty$ 
(high-density or weak-interaction limit) and to $V_{ee}^{\rm SCE}[\rho_\gamma]$ when $\gamma\to 0$ 
(low-density or strong-interaction limit). 

Standard KS DFT emphasizes the non-interacting shell structure, properly described through the 
functional $T_s[\rho]$, but it misses the features of strong correlation. SCE DFT, on the contrary, 
is biased towards localized ``Wigner-like'' structures in the density, accurately described 
by $V_{ee}^{\rm SCE}[\rho]$, missing the fermionic shell structure.
Many interesting systems lie in between the weakly and the strongly interacting limits, and 
their complex behavior arises precisely from the competition between the fermionic structure 
embodied in the kinetic energy and correlation effects due to the electron-electron repulsion. 
By implementing the exact $\tilde{v}_{\rm SCE}[\rho](\rv)$ potential in the Kohn-Sham scheme, we 
thus let these two factors compete in a self-consistent procedure.\cite{MalGor-PRL-12}

One should also notice that while the KS SCE approach does not use explicitly the Hartree 
functional, the correct electrostatics is still captured, since $V_{ee}^{\rm SCE}[\rho]$ is the 
classical electrostatic minimum in the given density $\rho$. Moreover, the potential 
$\tilde{v}_{\rm SCE}[\rho](\rv)$  stems from a wave function (the SCE 
one \cite{SeiGorSav-PRA-07,GorVigSei-JCTC-09}) and is therefore completely self-interaction 
free. 

Finally, another neat property of the zeroth-order KS SCE approach is that it always yields a 
lower bound to the exact ground-state energy $E_0=E[\rho_0]$, where $\rho_0$ is the exact 
ground-state density. In fact, for any given $\rho$ the right-hand side of Eq.~\eqref{eq_FKSSCE} 
is always less or equal than the left-hand side, as the minimum of a sum is always larger than 
the sum of the minima. As a consequence, for $\rho=\rho_0$ we have the inequality
\be
E[\rho_0]=F[\rho_0]+\int \rho_0\,v_{\rm ext}\ge T_s[\rho_0]+V_{ee}^{\rm SCE}[\rho_0]+\int \rho_0\,v_{\rm ext} \; ,
\label{eq_lowerbound}
\ee
which becomes even stronger when ones minimizes the functional on the right-hand-side with 
respect to the density within the self-consistent zeroth-order KS SCE procedure. It should be 
noted that this property implies an important difference with respect to the variational 
wave-function methods (such as HF, CI, QMC and DMRG), which, instead, provide an upper bound 
to the exact ground-state energy.
\subsection{Local correction to zeroth-order KS SCE}
\label{sec_KSSCELDA}
As preliminary found in Ref.~\onlinecite{MalGor-PRL-12} and further shown in Sec.~\ref{sec_results}, 
the zeroth-order KS SCE yields results that are qualitatively correct in the strong-correlation regime 
(representing a significative conceptual advance for KS DFT), but still with quantitative errors, 
which become smaller and smaller as correlation increases. An important issue is thus to add 
corrections to Eq.~\eqref{eq_FKSSCE}. One can, more generally, decompose $F[\rho]$ as
\be
F[\rho]=T_s[\rho]+V_{ee}^{\rm SCE}[\rho]+T_c[\rho]+V_{ee}^d[\rho],
\ee
where $T_c[\rho]$ (kinetic correlation energy) is 
\be
T_c[\rho]=\langle\Psi[\rho]|\hat{T}|\Psi[\rho]\rangle-T_s[\rho],
\ee
{\it i.e.}, the difference between the true kinetic energy 
and the Kohn-Sham one, and $V_{ee}^d[\rho]$ (electron-electron decorrelation energy) is 
\be
V_{ee}^d[\rho]=\langle\Psi[\rho]|\hat{V}_{ee}|\Psi[\rho]\rangle-V_{ee}^{\rm SCE}[\rho],
\ee
i.e., the difference 
between the true expectation of $\hat{V}_{ee}$ and the SCE value. A ``first-order'' 
approximation for $T_c[\rho]+V_{ee}^d[\rho]$ can be obtained from Eq.~\eqref{eq_EHxcSCE},
\be
T_c[\rho]+V_{ee}^d[\rho]\approx 2\,V_{ee}^{\rm ZPE}[\rho],
\ee
and can be, in principle, included exactly using the 
formalism developed in Ref.~\onlinecite{GorVigSei-JCTC-09}, but other approximations, {\it e.g.} 
in the spirit of Ref.~\onlinecite{SeiPerKur-PRA-00}, can also be constructed.  

Here we consider an even simpler approximation,  $T_c[\rho]+V_{ee}^d[\rho]\approx E_{\rm LC}[\rho]$, where  $E_{\rm LC}[\rho]$ is a local term that includes, at each point of space $\rv$, the corresponding correction for a uniform electron gas with the same local density $\rho(\rv)$, {\it i.e.},
\be
E_{\rm LC}[\rho]=\int\rho(\rv)\left[t_c(\rho(\rv))+v_{ee}^{d}(\rho(\rv))\right]d\rv.
\label{eq_LC}
\ee
In Eq.~\eqref{eq_LC} $t_c(\rho)$ and $v_{ee}^d(\rho)$ are the kinetic correlation energy and the electron-electron decorrelation 
energy per particle of an electron gas with uniform density $\rho$, corresponding to
\be
t_c(\rho)+v_{ee}^d(\rho)=\epsilon_{xc}(\rho)-\epsilon_{\rm SCE}(\rho),
\label{eq_corrSCE}
\ee
where $\epsilon_{xc}(\rho)$ is the usual electron-gas exchange-correlation energy and $\epsilon_{\rm SCE}(\rho)$ is the indirect part (expectation of $\hat{V}_{ee}$ minus the Hartree energy) of the SCE interaction energy per electron of the uniform electron gas with density $\rho$. This correction makes the approximate internal energy functional
\be
F[\rho]=T_s[\rho]+V_{ee}^{\rm SCE}[\rho]+E_{\rm LC}[\rho]
\ee
become exact in the limit of uniform density, similarly to what the LDA functional does in standard KS DFT.

\section{Model and Details of the Calculations}
\label{sec_Q1D}
We consider $N$ electrons in the quasi-one-dimensional (Q1D) model quantum wire of 
Refs.~\onlinecite{BedSzaChwAda-PRB-03,AbePolXiaTos-EJPB-07}, 
\be
\hat{H}=-\frac{1}{2}\sum_{i=1}^N \frac{\partial^2}{\partial x_i^2}+\sum_{i=1}^{N-1}\sum_{j=i+1}^N w_b(|x_i-x_j|)+\sum_{i=1}^N v_{\rm ext}(x_i),
\ee
in which the effective 
electron-electron interaction is obtained by integrating the Coulomb repulsion on the lateral 
degrees of freedom,\cite{BedSzaChwAda-PRB-03} and is given by 
\be
w_b(x)=\frac{\sqrt{\pi}}{2\,b}\,\exp\left(\frac{x^2}{4\,b^2}\right){\rm erfc}\left(\frac{x}{2\,b}\right).
\label{eq_int1D}
\ee
The parameter $b$ fixes the thickness of the wire, set to $b=0.1$ throughout this study, and ${\rm erfc}(x)$ is the complementary 
error function. The interaction $w_b(x)$ has a long-range coulombic tail, $w_b(x\to\infty)=1/x$, and is 
finite at the origin, where it has a cusp.
As in Ref.~\onlinecite{AbePolXiaTos-EJPB-07}, we consider an external harmonic confinement 
$v_{\rm ext}(x)=\frac{1}{2}\omega^2x^2$ in the direction of motion of the electrons.
The wire can be characterized by an effective confinement-length parameter $L$ such that 
$$\omega=\frac{4}{L^2},  \qquad v_{\rm ext}(x)=\frac{1}{2}\omega^2x^2.$$ 

\subsection{Zeroth-order KS SCE}
The co-motion functions $f_i(x)$ can be constructed by integrating Eqs.~(\ref{eq_fi}) for  
a given density $\rho(x)$,\cite{Sei-PRA-99,RasSeiGor-PRB-11,ButDepGor-PRA-12} choosing boundary conditions that make the density between two adjacent strictly-correlated positions always integrate to 1 (total suppression of fluctuations),\cite{Sei-PRA-99}
\be
\int_{f_i(x)}^{f_{i+1}(x)}\rho(x')\,dx'=1,
\ee
and ensuring that the $f_i(x)$ satisfy the required group properties.\cite{Sei-PRA-99,SeiGorSav-PRA-07,ButDepGor-PRA-12} This yields
\be
f_i(x)=
\Bigg\{
\begin{array}{l}
N_e^{-1}[N_e(x)+i-1] \qquad \qquad \; x\leq a_{N+1-i} \\
 N_e^{-1}[N_e(x)+i-1-N] \qquad x> a_{N+1-i},
\end{array} 
\ee
where the function $N_e(x)$ is defined as
\be
N_e(x)=\int_{-\infty}^x\rho(x')\,dx',
\ee
and $a_{k}=N_e^{-1}(k)$. Equation \eqref{eq_vSCE} becomes in this case
\be
\tilde{v}'_{\rm SCE}[\rho](x)=\sum_{i=2}^N w_b'(|x-f_i(x)|){\rm sgn}(x-f_i(x)).
\label{eq_vSCE1D}
\ee
We then solve self-consistently the Kohn-Sham equations \eqref{eq_KS}
with the KS potential $v_{\rm KS}(x)=v_{\rm ext}(x)+\tilde{v}_{\rm SCE}[\rho](x)$, 
where $\tilde{v}_{\rm SCE}[\rho](x)$ is obtained by integrating Eq.~\eqref{eq_vSCE1D} 
with the boundary condition $\tilde{v}_{\rm SCE}[\rho](|x|\to\infty)=0$. As said, we work in the spin-restricted KS framework, in which each spatial orbital is doubly occupied.

\subsection{The configuration interaction method (CI)}
In the configuration interaction calculations, the full many-body wavefunction is expanded as a linear combination of Slater determinants, constructed with the non-interacting harmonic oscillator orbitals. A matrix representation of the Hamiltonian in this basis is then numerically diagonalized to find the eigenstates of the system. The number of possible ways to place $N$ particles in a given set of orbitals increases rapidly as a function of $N$, such that only small particle numbers are tractable. Also, the stronger the interaction, the more basis orbitals are generally required to obtain a good approximation. For the present physical system, about 20--40 orbitals were needed to get converged solutions, which resulted in Hilbert space dimensions in the range $10^5$--$10^6$. For a more detailed description of the method, see e.g. Refs.~\onlinecite{RonCavBelGol-JCP-06,ReiMan-RMP-02}.

\subsection{KS LDA}
We have performed Kohn-Sham LDA calculations using the exchange-correlation
energy per particle $\epsilon_{xc}=\epsilon_x+\epsilon_c$ for a 1D homogeneous electron gas with the renormalized Coulomb 
interaction $w_b(x)$, as detailed in Ref.~\onlinecite{AbePolXiaTos-EJPB-07}. The exchange term $\epsilon_x$ is given by
\be
\epsilon_x(r_s)=\frac{1}{2}
\int_{-\infty}^{+\infty}\frac{dq}{2\pi}v_b(q)\left[S_0(q)-1\right]\; ,
\label{eq_exLDA}
\ee
where $v_b(q)$ is the Fourier transform of the interaction potential,
$S_0(q)$ is the non-interacting static structure factor, and
$r_s\equiv \frac{1}{2\rho}$.\cite{GiuVig-BOOK-05} To increase the numerical stability, we have interpolated between the Taylor expansions of $\epsilon_x(r_s)$ at small and large $r_s$ up to order 14.
For the correlation term we have used the results of Casula {\it et al.},\cite{CasSorSen-PRB-06} who have parametrized their QMC data as
\be
\epsilon_c(r_s)=-\frac{r_s}{A+B r_s^{\gamma_1}+C r_s^2}
\ln\left(1 + D r_s + E r_s^{\gamma_2}\right),
\label{eq_ecLDA}
\ee
where the different parameters are given in Table IV of 
Ref.~\onlinecite{CasSorSen-PRB-06} for several values of $b$.

\subsection{KS SCE with local correction}
\label{sec_KSSCELDA1D}
We have obtained the indirect SCE energy per electron $\epsilon_{\rm SCE}(\rho)$ needed in Eq.~\eqref{eq_corrSCE} by first computing the indirect $\epsilon^{\rm drop}_{\rm SCE}(\rho,N)$ for a 1D droplet with  $N$ electrons, uniform density $\rho$ and radius $R=\frac{N}{2\rho}$, as described in Ref.~\onlinecite{RasSeiGor-PRB-11}. We have then evaluated the limit $N\to\infty$ at fixed density $\rho$ to obtain the bulk value. The details of this calculation are reported in Appendix~\ref{app_SCELDA}. 

In Fig.~\ref{fig_eSCEgas} we show our numerical results for $b=0.1$ compared to the parametrized\cite{CasSorSen-PRB-06} QMC results for the exchange-correlation energy $\epsilon_{xc}(r_s)$ of Eqs.~\eqref{eq_exLDA}-\eqref{eq_ecLDA}. We see that, as it should be, $\epsilon_{\rm SCE}(r_s)\le \epsilon_{xc}(r_s)$ everywhere. For large $r_s$ we find that the SCE data are very close to the QMC parametrization, with differences of the order of $\sim 0.1\%$. Notice also that at $r_s=0$ we have $\epsilon_{\rm SCE}(0)=\epsilon_{xc}(0)=\epsilon_x(0)=-\frac{\sqrt{\pi}}{4 b}$. This is due to the fact that in the $r_s\to 0$ limit at fixed $b$ the first-order perturbation to the non-interacting gas is just a constant, so that every normalized wave-function yields the same result for the leading term. We have parametrized our data for $\epsilon_{\rm SCE}(r_s)$ as
\be
\epsilon_{\rm SCE}(\rho)=\rho\,q(2\,b\,\rho),
\label{eq_scalingeSCEunif}
\ee
with
\be
q(x)=A_1\,\ln\left(\frac{a_1 x+ a_2 x^2}{1+a_3 x+a_2 x^2}\right),
\label{eq_qfit}
\ee
and $A_1=0.9924534$, $a_2=1.55176743$, $a_3=2.025166778$, $a_1=a_3-\frac{a_2 \sqrt{\pi}}{2 A_1}$. This fit is valid for all values of $b$, since the scaling of Eq.~\eqref{eq_scalingeSCEunif} is exact for the SCE energy. The fitting function is also shown in Fig.~\ref{fig_eSCEgas} for the case $b=0.1$.
\begin{figure}
\includegraphics[width=8.5cm]{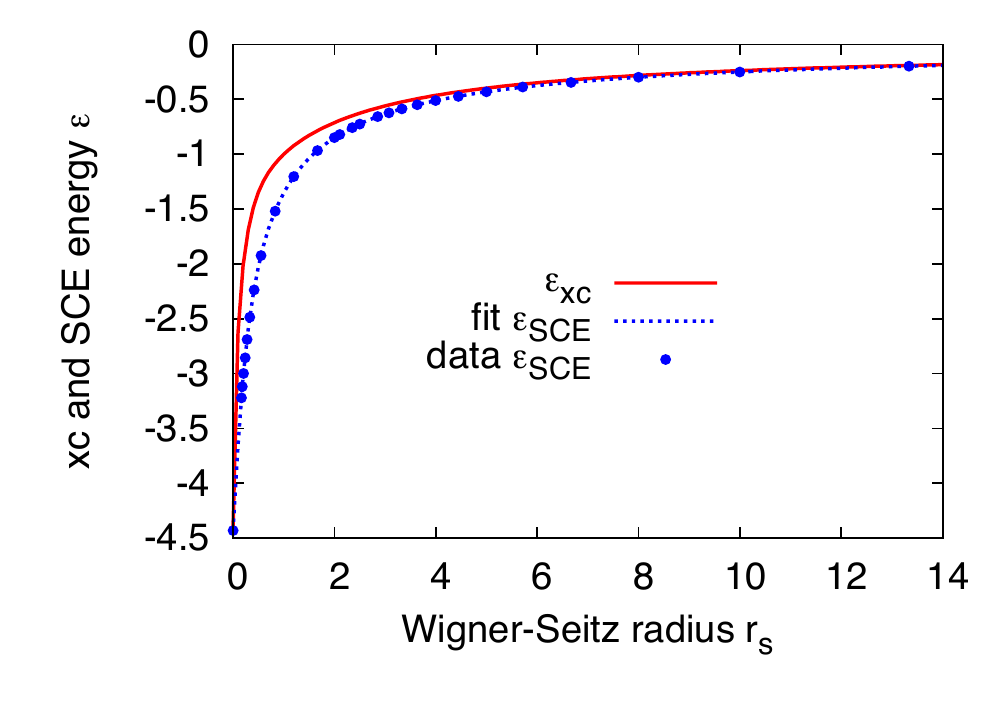}
\caption{(color online) The indirect SCE energy $\epsilon_{\rm SCE}(r_s)$ for the 1D gas [interaction of Eq.~\eqref{eq_int1D} and $b=0.1$] is compared to the parametrized QMC data\cite{CasSorSen-PRB-06} for the exchange-correlation energy $\epsilon_{xc}(r_s)$. For the SCE energy we show both our numerical results and the fitting function of Eqs.~\eqref{eq_scalingeSCEunif}-\eqref{eq_qfit}.} 
\label{fig_eSCEgas}
\end{figure}

\section{Results}
\label{sec_results}
\begin{figure}
\includegraphics[width=8.7cm]{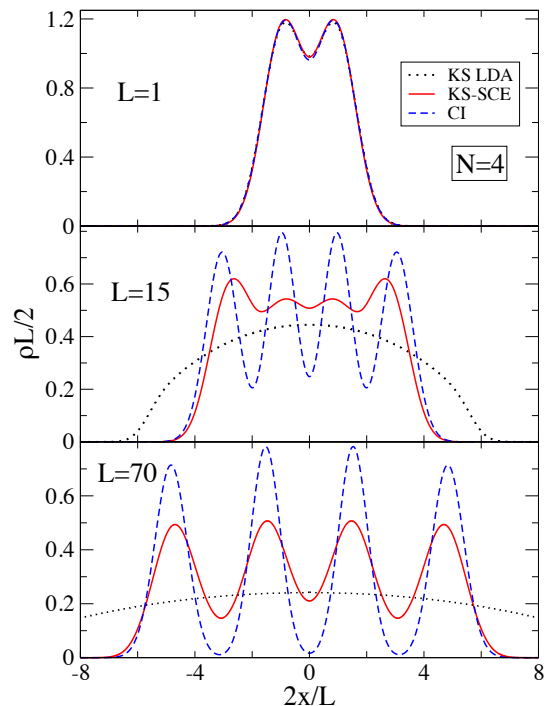}
\caption{(color online) Electron densities for $N=4$ and $L=1$, 15 and 70, obtained
with the KS SCE, CI and LDA approaches. The results are given in units of the 
effective confinement length $L=2 \omega^{-1/2}$.} 
\label{fig_densN6}
\end{figure}

Figure~\ref{fig_densN6} shows the electron densities for $N=4$ and different effective confinement lengths $L=2 \omega^{-1/2}$ obtained with 
the KS SCE, the CI and the KS LDA approaches. One can see that the three methods show qualitative agreement in the weakly-correlated regime, represented here in panel (a) by the 
case $L=1$. The densities have $N/2$ peaks, given by the Friedel-like oscillations with 
wave number 2$k_F^{\rm eff}$, where $k_F^{\rm eff} = \pi \tilde{\rho}/2$ is  the effective Fermi 
wavenumber, determined by the average density in the bulk of the trap $\tilde{\rho}$.

As the confinement length of the wire increases, the interactions start to become dominant 
and, whereas the KS SCE and the CI results are still in qualitative agreement, the 
LDA clearly provides a physical wrong description of the system. Indeed, one can see from 
panel (b) that whereas the densities obtained from the KS SCE and the CI methods 
develop a four-peak structure, corresponding to charge localization and 
indicating that the system enters the crossover between the weakly and the strongly 
correlated regimes (the $2 k_F\to 4 k_F$ crossover), the KS LDA yields a flat density. This is a typical error of local and semilocal density functionals that shows up also in bond breaking (yielding wrong molecular dissociation curves) and in systems close to the Mott insulating regime. In such cases, better total energies are obtained by using spin-dependent functionals and allowing symmetry breaking. This, however, does not yield a satisfactory physical description of such systems, missing many key features and giving a wrong characterization of several properties (see, e.g., Refs.~\onlinecite{AniZaaAnd-PRB-91,BorTorKosManAbeRei-IJQC-05,StoWagWhiBur-PRL-12}). 

When the system becomes even more strongly-correlated, here represented by $L=70$, the KS SCE gets closer to the
CI result, with densities that clearly present $N$ peaks, corresponding to charge localization. The KS LDA density is now very delocalized and almost flat in the scale of Fig.~\ref{fig_densN6}.
In order to obtain charge localization within the restricted KS scheme, the self-consistent KS potential must 
build ``bumps'' (or barriers) between the electrons. These barriers are a very non-local effect and are known to be a key property of the exact Kohn-Sham potential, as discussed in 
Refs.~\onlinecite{BuiBaeSni-PRA-89} and~\onlinecite{HelTokRub-JCP-09}.

\begin{figure}
\includegraphics[width=8cm]{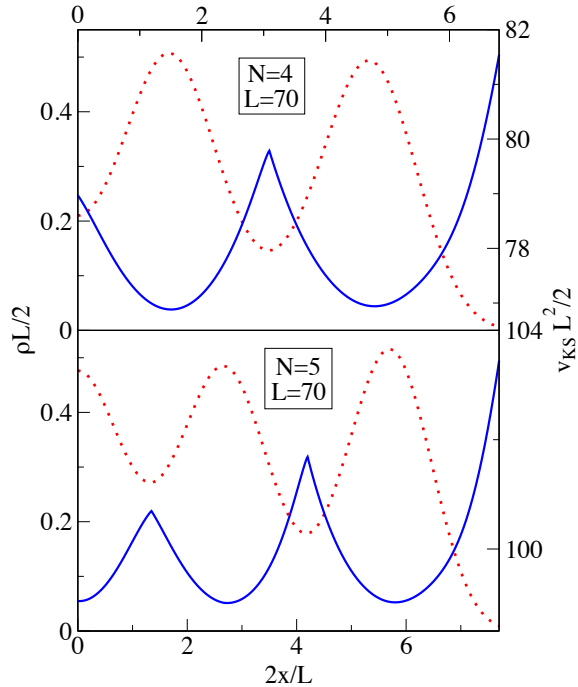}
\caption{(color online) Self-consistent Kohn-Sham potentials obtained with the KS-SCE method
for $N=4$ and $N=5$, with effective confinement length $L=70$ (blue solid lines). The corresponding densities
are also shown (red dotted lines). Notice that for the sake of clarity only the 
results for $x>0$ are shown. The results are given in units of the 
effective confinement length $L=2 \omega^{-1/2}$.} 
\label{fig_denspot1}
\end{figure}

In Fig.~\ref{fig_denspot1} we show that the self-consistent KS SCE scheme builds, indeed,
the above-mentioned barriers in the corresponding Kohn-Sham potentials, which
we plot together with the corresponding densities for $N=4$ and $N=5$ for $L=70$. 
One can see that each of the $N$ peaks in the density corresponds to a minimum 
in the KS potential, which is separated from the neighboring ones by 
barriers or ``bumps'', at whose maxima the KS potential has a discontinuous (but finite\cite{ButDepGor-PRA-12}) first
derivative. The number of such barriers is thus equal to $N-1$, and they become 
more pronounced with increasing correlation, enhancing the corresponding charge localization. Notice that the discontinuous first derivative of the KS SCE potential at the barrier maxima is a feature due to the classical nature of the SCE potential, and it is not expected to appear in the exact KS potential (indeed, it does not appear in any of the available calculations of the ``exact'' KS potential obtained by inversion).

\begin{figure}
\includegraphics[width=8.5cm]{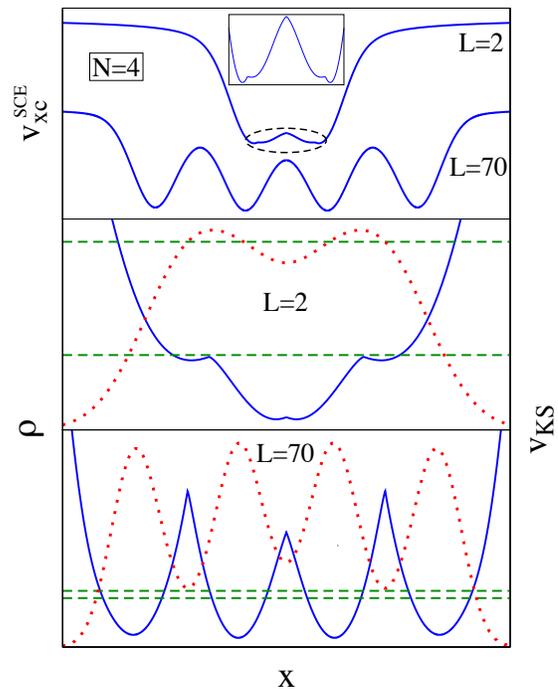}
\caption{(color online) Top panel: the self-consistent KS SCE exchange-correlation (xc) potential for $N=4$ at weak correlation ($L=2$) and strong correlation ($L=70$). In the inset, the oscillating part of the xc potential at $L=2$ is zoomed in. Middle panel: the total self-consistent KS SCE potential (blue, solid line), the corresponding density (red dotted line), and the two occupied KS eigenvalues (green dashed horizontal lines) for the weakly-correlated $L=2$ wire. In this case, we see that in the KS system there are no classically-forbidden regions inside the trap. Bottom panel: the same as in the middle panel for the strongly-correlated $L=70$ wire. In this case, we clearly see the classically-forbidden regions inside the trap created by the barriers in the KS SCE potential. The results are given in arbitrary units.} 
\label{fig_vxc_eigen}
\end{figure}

It is also interesting to make a connection between our results and the recent work on the KS exchange-correlation potential for the 1D Hubbard chains.\cite{VieCap-JCTC-10,Vie-PRB-12,Vie-arxiv-12} In particular, Vieira\cite{Vie-PRB-12} has shown that the exact exchange-correlation potential for a 1D Hubbard chain with hopping parameter $t$ and on-site interaction $U$, obtained by inversion from the exact many-body solution, always oscillates with frequency $4 k_F$, while the density oscillations undergo a $2 k_F\to 4k_F$ crossover with increasing $U/t$. The crossover in the density is thus due to the increase in the amplitude of the oscillations of the xc potential. In Fig.~\ref{fig_vxc_eigen} we show the KS SCE exchange-correlation potentials for $N=4$ electrons in the weakly ($L=2$) and strongly ($L=70$) correlated regimes. We see that the KS SCE self-consistent results are in qualitative agreement with the findings of Vieira:\cite{Vie-PRB-12} the oscillations in the xc potential have essentially a frequency $4 k_F$ also in the weakly-correlated case, with amplitude that increases with increasing $L$ [due to the scaling of Eqs.~\eqref{eq_scalTs}-\eqref{eq_scalF} the parameter $L$ plays here a role similar to $U/t$ for the Hubbard chain]. In the two lower panels of the same figure we also further clarify the $2 k_F\to 4k_F$ crossover in the KS framework: we see that the $4 k_F$ regime in the density oscillations occurs when the barriers in the total KS potential (due to the large oscillations of the xc potential) are large enough to create classically-forbidden regions inside the trap for the occupied KS orbitals. 

In Table~\ref{tab_totene} we report the total energies obtained with the three approaches, KS SCE, CI and KS LDA, for 
different values of the parameters $L$ and $N$. 
It can be seen that in the
weakly-correlated regime, represented here by $L=1$ and 2, the error made by
the KS SCE approach is larger than the one corresponding to the KS LDA. The results
also clearly show that, as previously discussed, KS SCE is always a lower bound to the total energy.
As the system becomes more correlated, the results obtained with the 
KS SCE and the CI approaches become closer to each other, whereas the
value given by the KS LDA is less accurate, as one could have inferred
from the corresponding densities shown in panels b) and c) of Fig.~\ref{fig_densN6}.

\begin{table}
\begin{tabular}{l*{5}{c}c}
\hline \hline
     $N$ &  $L$   &    & KS SCE & CI & KS LDA &\\
\hline
      2  &   2    &    & 1.81   &  2.49     &  2.59   &  \\
      2  &   15   &    & 0.0942 &  0.106    &  0.130  &  \\ 
      2  &   70   &    & 0.0112 &  0.0115   &  0.0182    &  \\
      4  &   1    &    & 25.08  &  28.42    &  28.57  &  \\
      4  &   2    &    & 8.46   &  10.60    &  10.68  &  \\ 
      4  &   15   &    & 0.491  &  0.541    &  0.580  &  \\
      4  &   70   &    & 0.0602 &  0.0629   &  0.0771    &  \\ 
      5  &   15   &    & 0.787  &  0.871    &  0.915  &  \\ 
      5  &   70   &    & 0.099  &  0.102    &  0.121    &  \\
\hline \hline
\end{tabular}
\caption{Comparison of the total energies obtained with 
the KS SCE, CI and KS LDA approaches for different values of
the particle number $N$ and effective-confinement length $L=2 \omega^{-1/2}$.}  
\label{tab_totene}
\end{table}

In the exact Kohn-Sham theory, the highest occupied KS eigenvalue is equal to 
minus the exact chemical potential from the electron-deficient side,\cite{AlmBar-PRB-85,LevPerSah-PRA-84} 
{\it i.e.}, $\mu^-=E_{N-1}-E_N$. 
In Table~\ref{tab_ioniz} we compare the highest occupied KS eigenvalue obtained with the KS SCE and the KS LDA 
approaches with the values of $E_{N}-E_{N-1}$ calculated from the total energies given by 
the CI method, corresponding to the same values of $N$ and $L$ given in Table~\ref{tab_totene}.
One can see that in this case the KS SCE gives good results also in the  
weakly-correlated regime.
In the strongly-correlated limit, the KS SCE and the CI results show an
agreement similar to that observed in the corresponding total energies. KS LDA, as usual, yields too high eigenvalues, due to the too fast decay of the exchange-correlation potential for $|x|\to\infty$.

\begin{table}
\begin{tabular}{l*{5}{c}r}
\hline \hline
     $N$ &  $L$   &    & KS SCE & CI & KS LDA &\\
\hline
      2  &   2    &    & 1.65   &   1.99  &  2.56   &  \\
      2  &   15   &    & 0.104  &  0.097  &  0.263  &  \\ 
      2  &   70   &    & 0.0126 &  0.0111 &  0.04087    &  \\
      4  &   1    &    & 11.26  &  11.86  &  12.56  &  \\
      4  &   2    &    & 4.08   &   4.65  &   5.02  &  \\ 
      4  &   15   &    & 0.248  &  0.256  &  0.453  &  \\
      4  &   70   &    & 0.0318 &  0.0304 &  0.06909    &  \\ 
      5  &   15   &    & 0.325  &  0.330  &  0.539  &  \\ 
      5  &   70   &    & 0.0408 &  0.0391 &  0.08172    &  \\
\hline \hline
\end{tabular}	
\caption{For the same systems of Table~\ref{tab_totene}, we compare the highest occupied KS eigenvalues obtained from KS SCE and KS LDA with the full CI values of $E_{N}-E_{N-1}$.}
\label{tab_ioniz}
\end{table}

\begin{figure}
\includegraphics[width=8.5cm]{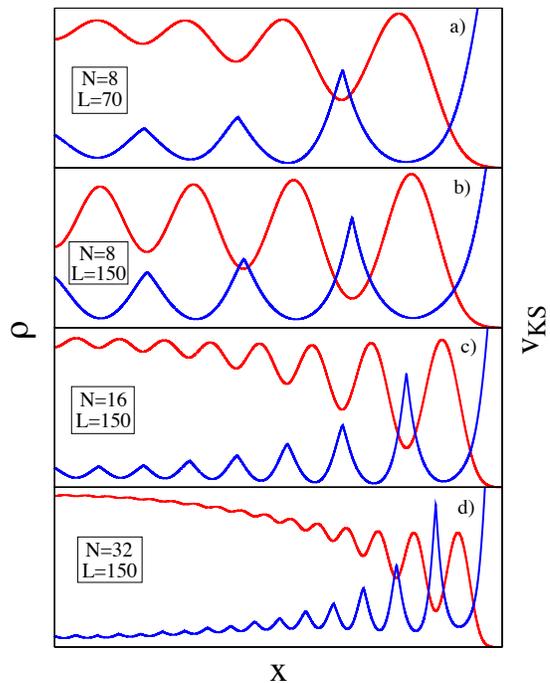}
\caption{(color online) Electron density and corresponding KS SCE
potential for different particle numbers $N$ and effective confinement lengths $L$. As in Fig.~\ref{fig_denspot1}, only the
results for $x>0$ are shown. The results are given in arbitrary units.} 
\label{fig_denspot2}
\end{figure}

As mentioned earlier, the numerical cost of the CI method increases 
exponentially with the number of particles, and this limitation 
becomes stronger as the correlations become dominant. In the 
calculations reported above, for the 5-electron case with $L=$70 we 
diagonalized a matrix where the eigenvectors had a dimension of about $3.5\times 10^5$. While it is technically possible to treat larger matrices, the rapid growth of the basis size still efficiently limits the number of particles one can handle. (For $N=6$ electrons, using the same basis orbitals, the corresponding dimension is roughly $2.6\times 10^6$.)
The KS SCE method, on the contrary, has a numerical cost (in 1D) comparable to the one of KS LDA,
therefore allowing to study strongly-correlated systems with much
larger particle numbers.
In Fig.~\ref{fig_denspot2} we show the electron densities and corresponding KS
potentials obtained with the KS SCE method for $N=8$, 16, and 32, for different values of $L$: in panels a) and b) we see how, at fixed number of particles $N=8$, the bumps in the KS potential and the amplitude of the density oscillations become larger with increasing $L$. For fixed effective confinement length $L=150$, we see from panels b), c) and d) how increasing the particle number $N$ leads to less pronounced features of strong correlation, according to the scaling of Eqs.~\eqref{eq_scalTs}-\eqref{eq_scalF}.

\begin{figure}
\includegraphics[width=8.5cm]{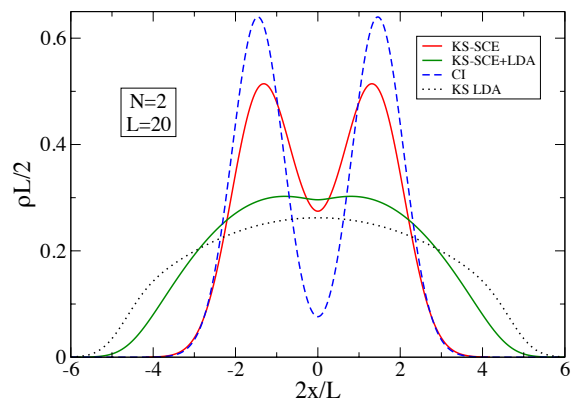}
\caption{(color online) Electron density for the case $N=2$ and $L=20$. The ``exact'' CI result is compared with the KS LDA, the KS SCE and the KS SCE with local correction of Secs.~\ref{sec_KSSCELDA} and \ref{sec_KSSCELDA1D} (KS SCE+LDA) results. The results are given in units of the 
effective confinement length $L=2 \omega^{-1/2}$.} 
\label{fig_KSCELC}
\end{figure}

Finally, we have tested the local correction to the zeroth-order KS SCE discussed in Secs.~\ref{sec_KSSCELDA} and \ref{sec_KSSCELDA1D}: as we see in the case $N=2$ and $L=20$ reported in Fig.~\ref{fig_KSCELC}, the results for the self-consistent densities are very disappointing, laying in between the KS SCE and the standard KS LDA values. This is due to the fact that, similarly to the standard KS LDA case, this simple local correction cannot capture the physics of the intermediate and strong-correlation regime, so that its inclusion worsens the results of KS SCE. In future work we will explore semi-local and fully non-local corrections to KS SCE.

\section{Conclusions and perspectives}
\label{sec_conc}
We have used the exact strong-interaction limit of the Hohenberg-Kohn 
functional to approximate the exchange-correlation energy and potential of Kohn-Sham 
DFT. By means of this so-called KS SCE approach, we have addressed 
quasi-one-dimensional quantum wires in the weak, intermediate and strong 
regime of correlations, comparing the results with those obtained by using 
the configuration interaction and the KS local density approximation.
In the weakly-correlated regime, the three approaches give qualitatively 
similar results, with electronic densities showing $N/2$ peaks, associated 
with the double occupancy of the single-particle levels that dominate the system. 
In this regime, KS LDA performs overall better than KS SCE.
As correlations become dominant, the KS SCE and the CI
densities start to develop additional maxima, corresponding to charge-density localization, whereas the KS LDA provides a 
qualitatively wrong description of the system, yielding a very
flat, delocalized, density.
We have also investigated a simple local correction to KS SCE, which, however, gives very disappointing results. In future works we will thus explore semi-local and fully non-local corrections to KS SCE.

The Kohn-Sham potential of the KS SCE approach shows ``bumps'' 
that are responsible for the charge localization and are a 
well-known feature of the exact Kohn-Sham potential of 
strongly-correlated systems. Moreover, the associated 
KS SCE exchange-correlation potential shows the right 
asymptotic behaviour, since it is self-interaction free
as it is constructed from a wave function (the SCE one\cite{SeiGorSav-PRA-07,GorVigSei-JCTC-09,MalGor-PRL-12}). This way, KS SCE is able to also give rather accurate chemical potentials. Notice that, as shown by studies of one-dimensional Hubbard chains, the $2k_F\to 4 k_F$ crossover in the density is a very challenging task for KS DFT for non-magnetic systems.\cite{VieCap-JCTC-10,Vie-PRB-12} The fact that KS SCE is able to capture this crossover is thus a very remarkable and promising feature.

Crucial for future applications is calculating $V_{ee}^{\rm SCE}[\rho]$ 
and $\tilde{v}_{\rm SCE}[\rho](\rv)$ also for general two- and three-dimensional systems. 
An enticing route towards this goal involves the mass-transportation-theory reformulation 
of the SCE functional,\cite{ButDepGor-PRA-12} in which $V_{ee}^{\rm SCE}[\rho]$ is given by 
the maximum of the Kantorovich dual problem,
\be
\max_{u}\left\{ \int u(\rv)\rho(\rv) d\rv \ : 
\ \sum_{i=1}^N u(\rv_i)\le \sum_{i=1}^{N-1}\sum_{j>i}^N \frac{1}{|\rv_i-\rv_j|} \right\}, \nonumber
\label{eq_VeeKantorovic}
\ee
where $u(\rv)=\tilde{v}_{\rm SCE}[\rho](\rv)+C$, and $C$ is a constant.\cite{ButDepGor-PRA-12} 
This is a maximization under linear constraints that yields in one shot the functional and its 
functional derivative. Although the number of linear constraints is infinite, this formulation may lead to  approximate but accurate approaches to the 
construction of $V_{ee}^{\rm SCE}[\rho]$ and $\tilde{v}_{\rm SCE}[\rho](\rv)$, as very recently shown by Mendl and Lin.\cite{MenLin-arxiv-12}

\section*{Acknowledgments}

We thank M. Seidl for inspiring discussions. This work was financially supported by the 
Netherlands Organization for Scientific Research (NWO) through a Vidi grant and by the Swedish Research Council and the Nanometer Structure Consortium at Lund University (nmC@LU).
\appendix
\section{SCE for the uniform Q1D electron gas}
\label{app_SCELDA}
Following Ref.~\onlinecite{RasSeiGor-PRB-11}, we have computed the SCE indirect Coulomb interaction energy per electron $\epsilon^{\rm drop}_{\rm SCE}(\rho,N)$ of a 1D droplet of uniform density $\rho$ and radius $R=\frac{N}{2\rho}$, where $N$ is the number of electrons,
\be
\epsilon^{\rm drop}_{\rm SCE}(\rho,N)=\frac{2}{N}\,\rho\, \tilde{v}_{ee}^{\rm SCE}(2\,\rho\, b,N)-\frac{\rho}{\pi}u_1\left(\frac{2\,b\,\rho}{N}\right),
\label{eq_eSCEdrop}
\ee
where
\be
\tilde{v}_{ee}^{\rm SCE}(x,N)=\frac{\pi}{2x}\sum_{i=1}^N(N-i)e^{i^2/x^2}{\rm erfc}\left(\frac{i}{x}\right)
\ee
is the rescaled SCE energy of the droplet\cite{RasSeiGor-PRB-11} and the second term in the right-hand-side of Eq.~\eqref{eq_eSCEdrop} is its Hartree energy, with
\be
u_1(x)=\int_0^\infty\left(\frac{\sin k}{k}\right)^2 e^{k^2 x^2}E_1(k^2 x^2)\, d k,
\ee 
and 
$$E_1(x)=\int_1^\infty\frac{e^{-t x}}{t}d t.  $$
Since the function $u_1(x)$ is numerically unstable, we have interpolated between its small-$x$ expansion through orders $O(x^5)$,
\be
	u_1^<(x)=\frac{\pi  x^4}{16}-\frac{\pi  x^2}{4}+\frac{1}{2}
	   \pi ^{3/2} x-\pi  \log (x)-\frac{1}{2} \pi  \psi
	   ^{(0)}\left(\frac{3}{2}\right),
	\nonumber
\ee
with $\psi^{(0)}\left(\frac{3}{2}\right)\approx 0.036489974$, and its large-$x$ expansion through orders $O(x^{-16})$,
\begin{eqnarray*}
& &	u_1^>(x)=\frac{\pi ^{3/2}}{1209600 x^{15}}-\frac{64 \pi
	   }{14189175 x^{14}}+\frac{\pi ^{3/2}}{131040
	   x^{13}}- \\
& &	\frac{16 \pi }{405405 x^{12}}+\frac{\pi
	   ^{3/2}}{15840 x^{11}}-\frac{16 \pi }{51975
	   x^{10}}+\frac{\pi ^{3/2}}{2160 x^9}-\frac{2 \pi
	   }{945 x^8} \\
	& & +\frac{\pi ^{3/2}}{336 x^7}-\frac{4
	   \pi }{315 x^6}+\frac{\pi ^{3/2}}{60
	   x^5}-\frac{\pi }{15 x^4}+\frac{\pi ^{3/2}}{12
	   x^3}-\frac{\pi }{3 x^2}+\frac{\pi ^{3/2}}{2 x},
\end{eqnarray*}
switching between them at $x=0.584756$.

We have then evaluated numerically the limit $N\to\infty$ of Eq.~\eqref{eq_eSCEdrop} at fixed $\rho$. We have found that the convergence is reasonably fast: for example, taking $N=10^5$ yields results with a relative accuracy of $10^{-6}$. Our numerical results have been fitted with the function $q(x)$ of Eqs.~\eqref{eq_scalingeSCEunif}-\eqref{eq_qfit}.
 

\end{document}